\input harvmac 
\input epsf.tex
\overfullrule=0mm
\newcount\figno
\figno=0
\def\fig#1#2#3{
\par\begingroup\parindent=0pt\leftskip=1cm\rightskip=1cm\parindent=0pt
\baselineskip=11pt
\global\advance\figno by 1
\midinsert
\epsfxsize=#3
\centerline{\epsfbox{#2}}
\vskip 12pt
{\bf Fig. \the\figno:} #1\par
\endinsert\endgroup\par
}
\def\figlabel#1{\xdef#1{\the\figno}}
\def\encadremath#1{\vbox{\hrule\hbox{\vrule\kern8pt\vbox{\kern8pt
\hbox{$\displaystyle #1$}\kern8pt}
\kern8pt\vrule}\hrule}}


\def\mod{{\rm mod \ }}
\font\cmss=cmss10 \font\cmsss=cmss10 at 7pt
\def\IZ{\relax\ifmmode\mathchoice
{\hbox{\cmss Z\kern-.4em Z}}{\hbox{\cmss Z\kern-.4em Z}}
{\lower.9pt\hbox{\cmsss Z\kern-.4em Z}}
{\lower1.2pt\hbox{\cmsss Z\kern-.4em Z}}\else{\cmss Z\kern-.4em Z}\fi}

\Title{SPhT/94-018; cond-mat/9402058 }
{{\vbox {
\bigskip
\centerline{Entropy of Folding of the Triangular Lattice}
}}}
\bigskip
\centerline{P. Di Francesco}
\medskip
\centerline{and} 
\medskip
\centerline{E. Guitter,}

\bigskip

\centerline{ \it Service de Physique Th\'eorique de Saclay
\footnote*{Laboratoire de la Direction des Sciences 
de la Mati\`ere du Commissariat \`a l'Energie Atomique.},}
\centerline{ \it F-91191 Gif sur Yvette Cedex, France}

\vskip .5in
The problem of counting the different ways of folding 
the planar triangular lattice is shown to be equivalent to 
that of counting the possible $3$--colorings of
its bonds, a dual version of the $3$--coloring problem of the 
hexagonal lattice solved by Baxter. The folding entropy $\hbox{Log} \ q$
per triangle is thus given by Baxter's formula 
$q=\sqrt{3}\ {\Gamma(1/3)^{3/2}/ 2\pi}=1.2087... $ .

\noindent
\Date{02/94}

\nref\NP{D.R. Nelson and L. Peliti, J. Phys. France {\bf 48} (1987) 1085.}
\nref\KN{Y. Kantor and D.R. Nelson, Phys. Rev. Lett. {\bf 58} (1987) 2774
and Phys. Rev.  {\bf A 36} (1987) 4020.}
\nref\PKM{M. Paczuski, M. Kardar and D.R. Nelson, Phys. Rev. Lett. {\bf 60} (1988) 2638.}
\nref\DG{F. David and E. Guitter, Europhys. Lett. {\bf 5} (1988) 709.}
\nref\KAN{Y. Kantor and M.V. Jari\'c, Europhys. Lett. {\bf 11} (1990) 157.}
\nref\BT{R.J. Baxter and S.K. Tsang, J. Phys. {\bf A13} Math. Gen. (1980) 1023.} 
\nref\BAX{R.J. Baxter, J. Math. Phys. {\bf 11} (1970) 784 and 
J. Phys. {\bf A19} Math. Gen. (1986) 2821.}


The link between geometrical objects like polymers or
membranes and spin systems is a powerful tool of investigation.
The case of phantom polymers is especially eloquent, since in its discrete
form a phantom chain can be thought of as
$1$--dimensional Heisenberg chain, in which the spin variables stand for
the directions of elementary monomers.  The Heisenberg coupling 
is nothing but a bending energy for the (semi--flexible) polymer. 
Analogously, the folding of a chain is expressible as a
$1$--dimensional Ising model.

The $2$--dimensional case is somewhat more subtle.
Generically, $2$--dimensional membranes can be discretized into
triangulations, whose faces can support
Heisenberg spin variables, now representing the direction of the local
normal vector to the surface.
However, one finds several universality
classes of phantom membranes, depending on internal order and/or rigidity \NP.
Beside the case of fluid membranes, which correspond to random triangulations, 
a case of special interest is that of tethered or polymerized membranes,
which correspond to triangulations with fixed connectivity. 
In their simplest formulation, tethered membranes translate into
a statistical problem on the regular triangular lattice \KN. 
The corresponding spin system is more involved than a simple $2$--dimensional
triangular Heisenberg model: indeed, as normals to a surface, 
the spins are not independent variables.  
The resulting constraints play a crucial role in favorizing an ordered
phase and are responsible for the existence of a crumpling
transition for tethered membranes [\xref\NP -\xref\DG ], 
as opposed to the case of the usual (unconstrained)
$2$--dimensional Heisenberg model, always disordered.

In this letter, we address the similar, but simpler
question of folding of the regular
triangular lattice, the geometrical counterpart of a triangular
$2$--dimensional Ising model modified by additional local constraints.
This model has been introduced and studied numerically in ref.\KAN,
as a simple example in which connections
between discrete and continuous models can be analyzed.
It also appears to be useful for discussing
analogies and differences between spin
systems and geometrical objects.

After defining the folding problem and its spin formulation, we establish
an equivalence between this model and the $3$--coloring 
problem of the bonds of the triangular lattice. 
As a main result,
this allows us to obtain the exact value of the folding entropy per triangle,
which counts the different ways of folding the triangular lattice.

\bigskip

{\bf The Model.}
We define a folding as a mapping
which assigns to each vertex of the triangular lattice a position in the plane,
with the ``metric constraint" that the distance in the plane between nearest 
neighbors on the lattice is maintained equal to, say, unity. 
Under such a folding, each elementary triangle of the lattice
is mapped onto an equilateral triangle of the plane and each bond of 
the lattice serves as a hinge between its two neighboring triangles, and
can be folded or not, depending on their relative
position in the folded state.
Up to
global translations and rotations, a folded state is actually
uniquely specified 
by the data of these elementary folds along lattice bonds. 
 
\fig{The eleven local fold environments for a vertex. Folds are represented by
thick lines.}
{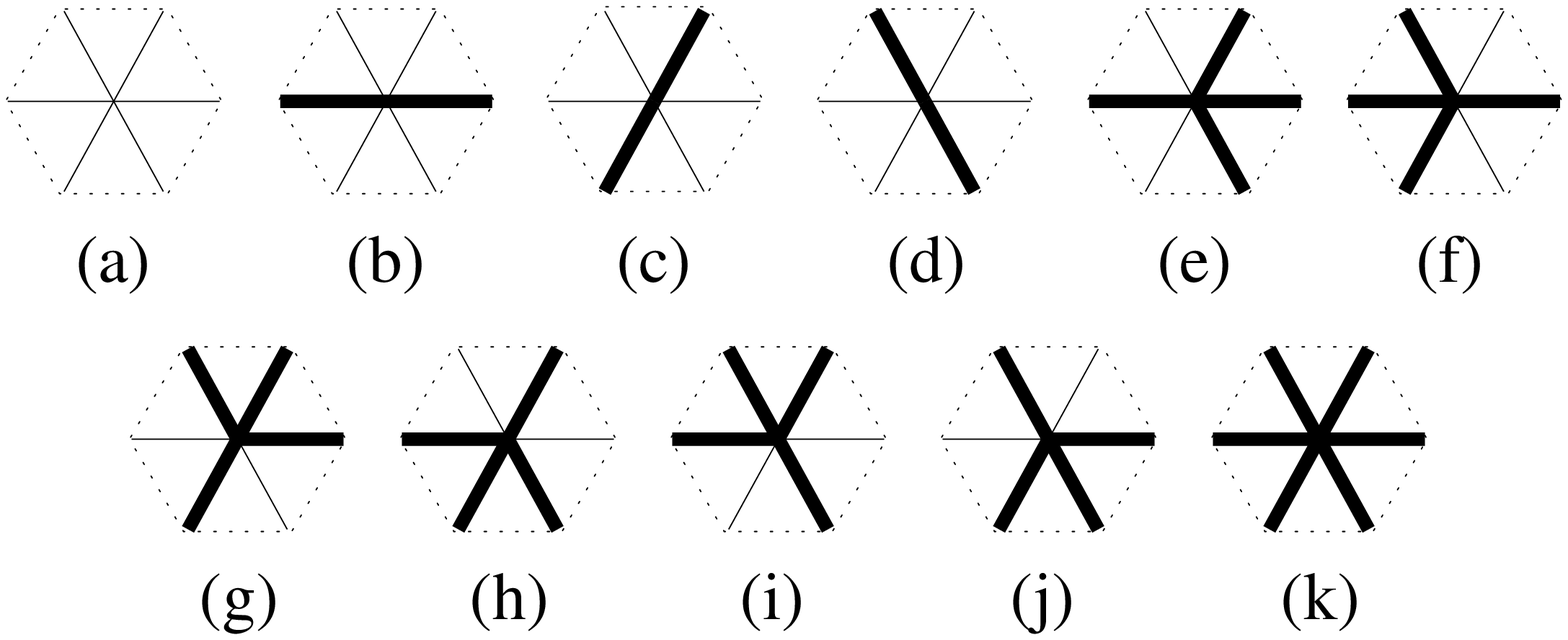}{8. truecm}
\figlabel\eleven

\noindent
It is easy to verify that, among the $2^6$ possible fold configurations 
for the six bonds surrounding a given vertex,
only $11$ are allowed, corresponding to actual local foldings of the hexagon
surrounding the vertex. 
These configurations are depicted on Fig.\eleven : configuration
(a) corresponds to no fold, (b-d) to one fold of the hexagon made of
{\it two} elementary bond folds, 
(e-j) to two folds of the hexagon ({\it four} elementary bond folds)
and (k) to three folds of the hexagon ({\it six} elementary bond folds).  
Note that, with the above definition, a folding does not
distinguish between the different ways of folding leading
to the same final state: by distinct foldings, we actually mean 
distinct folded state.
In its original version \KAN, the folding problem can therefore
be expressed as an eleven vertex model.
Natural weights for the vertices of Fig.\eleven\ are
$w_a=1$, $w_{b-d}=z$, $w_{e-j}=z^2$, $w_k=z^3$, which amounts to
assign the weight $z$ per elementary fold (the power of $z$ in the
weights $w$ is half the number of elementary folds, 
since a bond shared by two vertices is counted twice).
Since all vertices on Fig.\eleven\ have an even number of elementary folds,
these folds can be organized into folding lines without endpoints.

One can also think of foldings in terms of spin variables $\sigma_i=\pm$
living on the triangles, which indicate whether the $i$--th triangle 
faces up or down in the folded state. 
The spin value changes between
two neighboring triangles if and only if their common bond lies on a 
folding line, i.e. folding lines are domain walls of the spin system.
This actually leads to two possible spin configurations per folded
state, due to the degeneracy under reversal of all spins.
\fig{Elementary folding configurations and their corresponding
$\pm$ spin configurations, up to global reversal of all spins.}{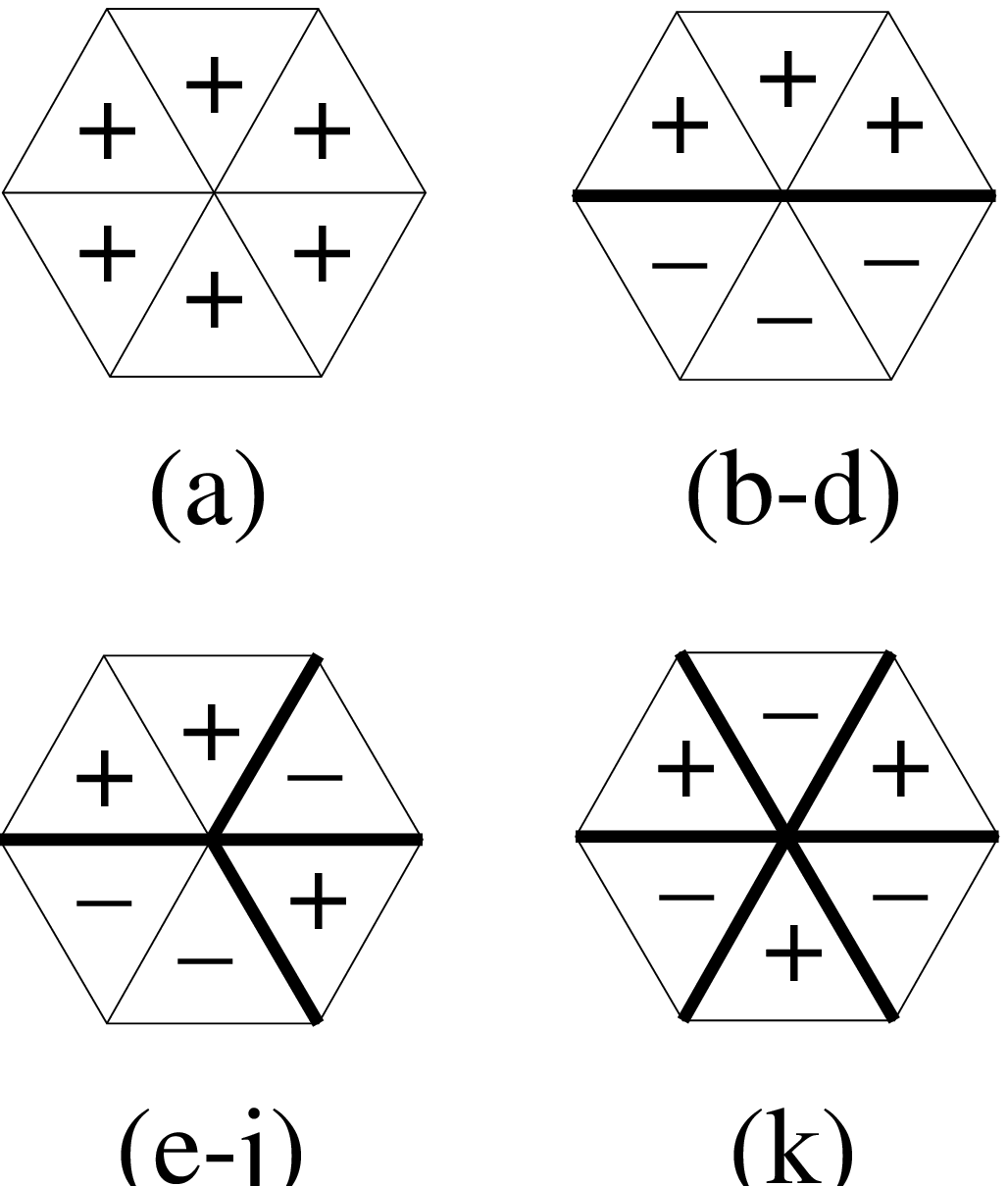}{6. truecm}
\figlabel\fold
\noindent
Conversely, arbitrary spin configurations will not in general arise from
folded states.
By examining the $11$ possible elementary foldings (see Fig.\fold ),
we obtain a local constraint on the spin configurations around each vertex:
the number of $+$ spins around a vertex must be
a multiple of $3$ (i.e. $0$, $3$ or $6$). 
This local constraint turns out to
be sufficient to ensure that the spin configuration actually corresponds 
to a folded state. The number of constrained spin configurations
is indeed twice that of foldings of the hexagon (${ 6 \choose 0}+
{ 6 \choose 3}+{ 6 \choose 6}=2 \times 11$), as wanted.
We see here, as mentioned in the introduction, 
an explicit realization of constraints distinguishing the geometrical 
folding problem from an ordinary Ising model, and ensuring that the
spin variables are in fact normals to a (tethered) surface.

\noindent
In this language, the above statistical weight $z$ per folded bond
translates into an Ising--like interaction between nearest 
neighboring spins, with coupling constant 
$\beta J = - {1 \over 2} \hbox{Log} \ z$.
In the following we will restrict ourselves to $z=1$ and to the
problem of counting the folding configurations.

\bigskip

{\bf Entropy.}
Let us consider a finite lattice with $N$ elementary triangles, and 
denote by $Z_N$ the number of its possible foldings. 
The entropy of folding
per lattice face is defined as the thermodynamical limit
$$s=\lim_{N \to +\infty} {1 \over N} \hbox{Log}\ Z_N = \hbox{Log} \ q \ .$$
In the spin language (where the number of spin configurations is clearly
$Z_{\rm spin}= 2 Z_N$), $q$ is simply the average number of 
spin configurations per triangle, and in particular $1 \leq q \leq 2$.
Such a behavior $Z_N \propto q^N$ with a non-zero value of $s$ (i.e. 
$q>1$) has been 
supported numerically in \KAN, 
where it is found that $q \simeq 1.21$, by use of a transfer matrix 
formalism for the above $11$--vertex model.

The non--vanishing of $s$ is not intuitively obvious.
Indeed, starting from the unfolded state with all spins $+$, considered as a
ground state, there are {\it no local excitations}, i.e. the reversal of one spin
implies that of a line of them all the way to the boundary.
The reason for that is best seen on the $11$--vertex version, since
by drawing a vertical line across each vertex  
of Fig.\eleven, one notices that whenever a fold is present
on the left side of this line (cases (b--k)), 
at least another one continues to the right, and conversely.
Therefore, any elementary fold is part of a line crossing the entire
lattice from the left to the right.
In the case of the folding problem of the square lattice \DG, an analogous
phenomenon of non--locality
results in the vanishing of the entropy $s_{\rm square}$
(the number of folded states grows like $4^{L}$  for a
square lattice of size $L \times L$, with $N=L^2$ faces, hence
$s_{\rm square}(L)={1 \over L^2} \hbox{Log}\ 4^L 
{{\ } \atop {\displaystyle\longrightarrow \atop {L\to \infty}}}  0$).

It is therefore interesting to directly prove the non--vanishing of $s$.
This can be done by changing the definition of our ground state, namely
choosing now the completely folded state, for which
we pick the ``antiferromagnetic"
spin configuration with $+$ on all triangles pointing up,
and $-$ on the others.
\fig{Hard hexagon excitations: the spins inside the shaded hexagons 
are reversed with respect to the antiferromagnetic background.}{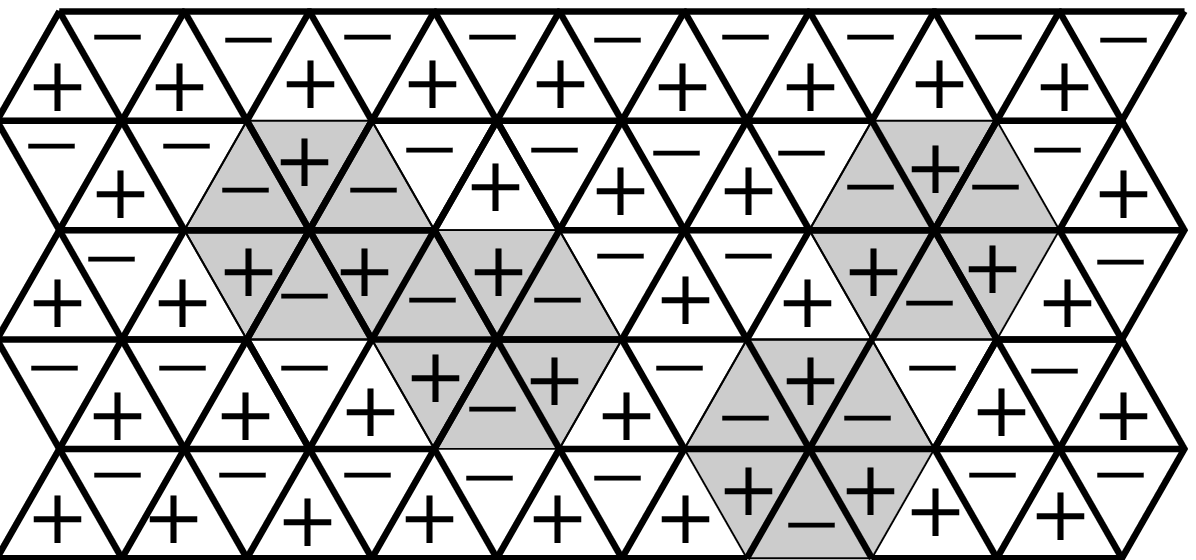}{9. truecm}
\figlabel\hexa
\noindent {\it Local excitations}
now exist, which correspond to reversing the six
spins inside an elementary hexagon (see Fig.\hexa\ ). 
Indeed the creation of such an hexagon only affects its central and
peripheral vertices, for which it exchanges spins by pairs of a $+$
and a $-$ spin, leaving
the number of $+$ spins surrounding these vertices unchanged.
A gas of such
excitations is also allowed, provided that two reversed hexagons do
not overlap (see Fig.\hexa\ ): this is precisely the rule for the celebrated
hard hexagon model, which obviously has a non--zero entropy per triangle 
$s_{\rm hh}>{1\over 6}\ \hbox{Log}\ 2$. 
This first shows the non--vanishing of the entropy of folding $s$ and moreover
provides a rough lower bound $q \ge 2^{1/6}$. 
Baxter and Tsang \BT\ obtained
the numerical estimate $2s_{\rm hh}=.33324$, which gives a much
better bound $q> 1.1813$. 
Comparison with the numerical estimate of \KAN\ shows
that hard hexagon excitations contribute to more than $85$ percent of
the entropy of folding.

\bigskip

{\bf Equivalence with $3$--colorings.}
Our main point in this letter is to establish that the folding problem
is equivalent to a $3$--coloring problem of the bonds of the 
triangular lattice. 

We use three colors, blue (B), white (W) and red (R), ordered 
cyclically so that red follows white follows blue follows red.
The colors can also be thought of as elements of $\IZ_3=\{0,1,2\}$, defined
modulo $3$.
\fig{Color assignments for the lattice bonds: for a $+$ (resp. $-$) spin,
the colors increase (decrease) counterclockwise.}{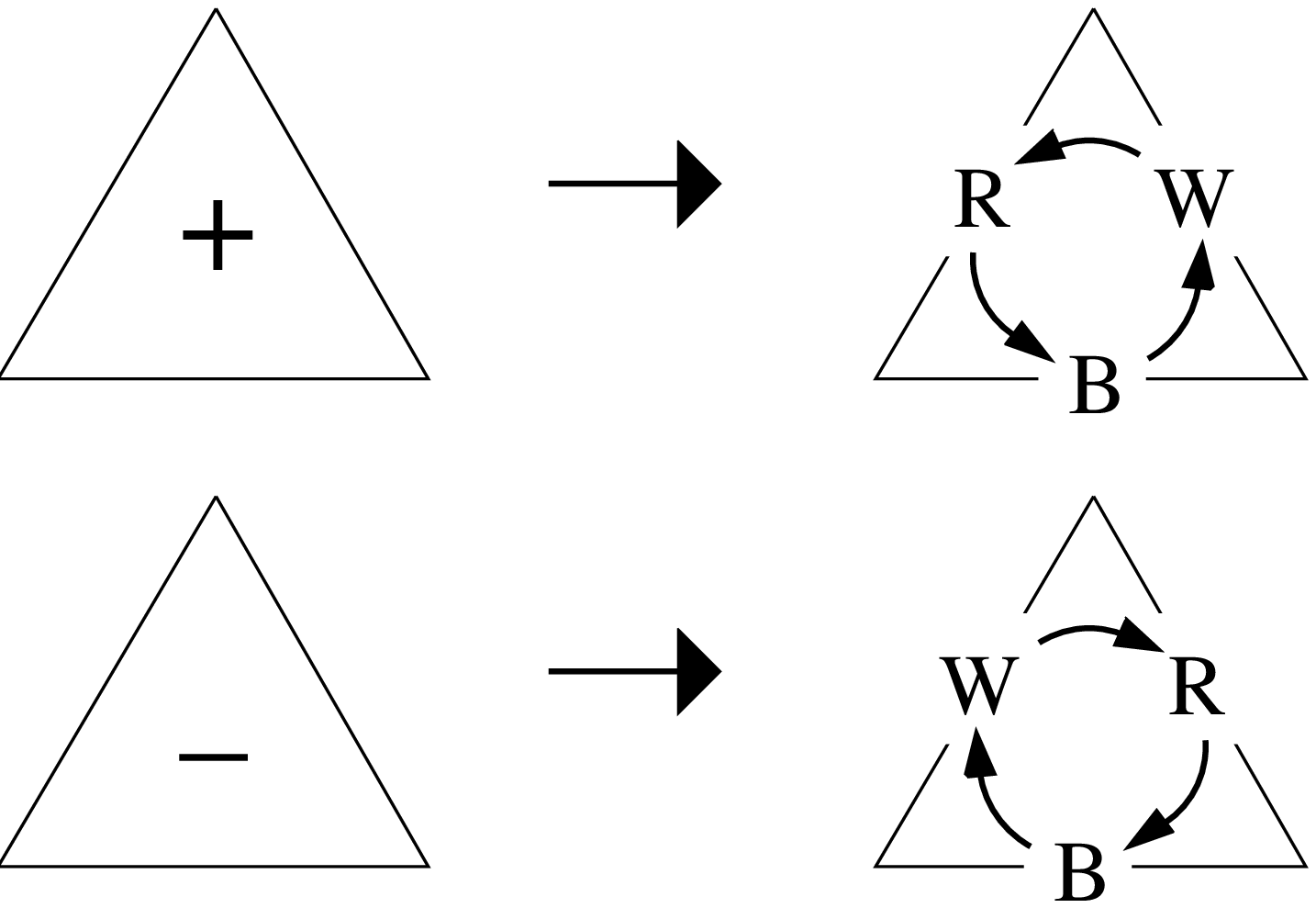}{7. truecm}
\figlabel\colspin 
Starting from a spin configuration, we pick a bond to which we assign 
a color, say blue, and paint step by step all the bonds
with the rule that the three bonds of any triangle have distinct colors
B, W, R, which are chosen in increasing (resp. decreasing)
order counterclockwise, if the spin is $+$ (resp. $-$), as in Fig.\colspin. 
This rule fixes the colors of the bonds unambiguously, thanks to the
local constraint on the spins. 
\fig{The relation $c_{i+1}=c_i+\sigma_i\ \mod 3$ between spin and color
variables around a vertex. Coloring is unambiguous if $c_7=c_1$.}
{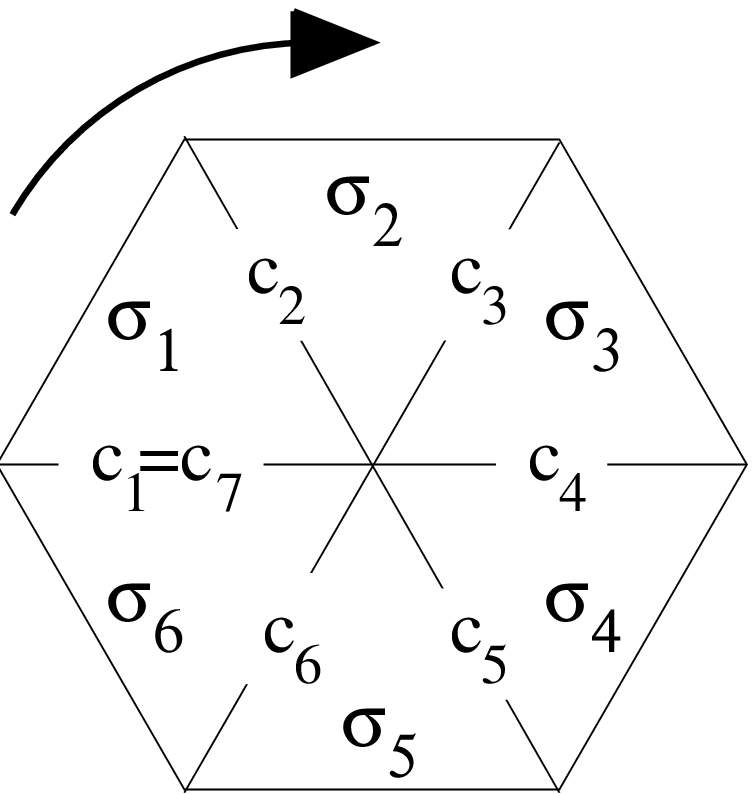}{5. truecm}
\figlabel\spincol 
To see this,
it is sufficient to examine the six bonds around a vertex 
(see Fig.\spincol\ ):   
following these bonds clockwise, we see that, according to our rule,
the color $c_{i+1}$ of the $(i+1)$--th bond 
is obtained by adding to the color $c_i$ of the $i$--th bond the
algebraic value $\pm$ of the crossed spin $\sigma_i$:
$c_{i+1}=c_i+\sigma_i \ \mod 3$.
The coloring is well defined iff, after a complete turn, 
$c_7=c_1 \ \mod 3$, which amounts to
$\Sigma=\sigma_1+\sigma_2+\sigma_3+\sigma_4+\sigma_5+\sigma_6=0 \ \mod 3$.
On the other hand, $\Sigma = 2 \times \hbox{(number of $+$)} -6$,
hence $\Sigma$ is a multiple of $3$ iff the number of $+$ spins around
the vertex is a multiple of $3$, which is precisely our folding
constraint on the spins.

To each spin configuration we have attached a $3$--coloring of the
bonds such that {\it the three colors around each 
triangle are distinct}. Conversely any such coloring leads to a
spin configuration (spin value $+$ if the colors increase counterclockwise,
$-$ otherwise)
satisfying the folding constraint. The correspondence
is $1$ to $3$, due to the $3$ possible choices for the color of the first bond.
The number of $3$--colorings is $Z_{\rm color}=3Z_{\rm spin}=6Z_N$.

It is interesting to notice that, in the original folding
problem, the colors have a natural interpretation
as the orientation of the bonds in the folded state. 
More precisely, let us fix a direction for the bonds of the triangular 
lattice, such that all the horizontal bonds point to the right, all
bonds rotated by $2 \pi /3$ point up, and all bonds rotated by 
$4 \pi /3$ point down.
The image of a directed bond $i$ in the folded state is 
a unit vector $\vec t_i$, which can only take three values: a unit
vector
$\vec e_B$ which fixes the (arbitrary) overall orientation of the folded 
state, its image $\vec e_W$ by a rotation of $2 \pi /3$, or
its image $\vec e_R$ by a rotation of $4 \pi /3$ (with of course
$\vec e_B + \vec e_W + \vec e_R =\vec 0$).
One can easily convince oneself that the actual value of $\vec t_i$
is simply fixed by the color $c_i$ of the bond $i$: $\vec t_i=\vec e_{c_i}$.
In this respect, the $3$--coloring model simply reexpresses the folding
problem in terms of tangent vectors $\vec t_i$
to the surface rather than normals.

The $3$--coloring problem introduced above was exactly solved by Baxter 
\BAX\ in its equivalent dual form (i.e. coloring the bonds of the
hexagonal lattice with three colors so that no concurrent bonds 
are colored alike).
Using a transfer matrix method and Bethe ansatz techniques, Baxter obtains
the following result for the entropy of $3$--coloring,
now reinterpreted as the {\it exact entropy of folding} $s=\hbox{Log}\ q$:
$$ q= \prod_{n=1}^{\infty} { (3n-1) \over \sqrt{ 3n(3n-2)} }=
{\sqrt{\Gamma(1/3)} \over \Gamma(2/3) }= {\sqrt{3} \over 2 \pi}
\Gamma(1/3)^{3/2}.$$
Its numerical value $q=1.208717...$ ($s=.189560...$) is in very good 
agreement with the numerical estimate of \KAN.

\bigskip

In the above, we established the equivalence between the folding problem of the 
triangular lattice and the exactly solvable $3$--coloring problem of its bonds. 
This fact sheds a new light on the initial geometrical problem, and offers 
perspectives for further studies.

A first direction concerns the evaluation of correlation functions for 
geometrical quantities in the folding problem. For instance the
correlation between normal vectors to the folded surface is exactly
that of the spin variables, also expressible in terms of correlations
of colors.  Another quantity of interest is the mean square
distance between two points in the folded state. We choose for
instance two points $P$ and $Q$
on the same direction of the lattice, distant by $d$
lattice spacings. Using the interpretation of colors as orientations
of the bonds in the folded state, the square distance between P and Q
in the folded state reads
$$r_{\rm PQ}^2 = N_B^2 + N_W^2+ N_R^2 -N_B N_W - N_W N_R - N_R N_B ,$$ 
where $N_{\rm B,W,R}$ denote respectively the numbers of B,W,R--colored
bonds on the segment joining $P$ and $Q$.
Based on numerical simulations \KAN, we expect 
$\langle r_{\rm PQ}^2 \rangle \propto \hbox{Log}\ d$, the usual behavior
for phantom polymerized membranes. This would amount to a color--color
correlation which decreases as ${1 / d^2}$.

A second direction of study is the understanding of the effect of rigidity by
introducing a weight $z=\exp -2\beta J <1$ per fold, with the issue of a
possible crumpling transition. Numerical results \KAN\ seem to indicate that a
crumpling transition takes place at $z=1$, i.e. for an infinite temperature
($\beta =0$), with the system being in an ordered phase for any finite 
temperature. This agrees with the idea that a folding constraint on the
spins indeed favorizes an ordered phase. 
In this letter, we showed that without rigidity but with the folding 
constraint, 
the model turns out to be exactly solvable.
With rigidity, but no folding constraint, the model is just the triangular
Ising model, exactly solvable too. 
These are encouraging indications that an exact
solution may be found for the rigid folding problem.

\medskip

We thank F. David, O. Golinelli and J.--M. Luck for useful discussions.

\listrefs

\bye